\documentclass[showkeys,showpacs,amsmath,superscriptaddress,preprintnumbers,amssymb,10pt,twocolumn]{revtex4}
\usepackage{graphicx}
\usepackage{bm}
\usepackage{color}
\usepackage{subfigure}
\usepackage{amssymb}

\begin{document}
\title{On the information entropy of matter-waves in quasi-disorder potentials}
\author{Kajal Krishna Dey}
\affiliation{Department of Physics (UG $\&$ PG), Banwarilal Bhalotia College,  Asansol-713303, India}
\author {Sudipta Das}
\affiliation{Department of Physics, Government General Degree College, Chapra Shikra, Nadia-741123, India}
\author{Golam Ali Sekh}
\email{skgolamali@gmail.com}
\affiliation{Department of Physics (UG $\&$ PG), Banwarilal Bhalotia College,  Asansol-713303, India}

\begin{abstract}
We consider ultracold Bose gases in quasi-random  potentials and quantify localization of matter waves  by means of Shannon information entropy.  We explicitly examine the role of quasi-random potentials in producing localized states in the linear and nonlinear regimes. It is seen that the information entropic-based approach  can be more useful to  quantify localization of different types of states observed in  Bose-Einstein condensates. 
\end{abstract}
 \maketitle
\section{Introduction}
An important problem in the context of many-body quantum systems is  understanding the effects of interactions. Ultracold bosonic atoms  have been proved to be a powerful tool to reproduce several interacting physical systems with great flexibility over different parameters \cite{books1,book3,book4,book5,Dalfavo,Leggett,Pitaevskii}. A disorder potential in the ultra-cold atomic system can naturally be induced either from  imperfection in magnetic-trap wire \cite{r1,r1a,r2,r2aa,r2a} or during the fabrication  of device \cite{r3}. In the presence of disorder potentials the interacting bosons leads to the so-called {\it dirty boson problems}. It offers a platform to understand the interplay between the disorder potential and the mean-field interaction in producing localized and super-fluid states. Understanding the interplay between localized and super-fluid phases is a long standing problem in condensed matter physics \cite{r3a,r3b}.

In the negligible interaction limit, the weak disorder leads to exponential localized one-body wave-function,  often called the  Anderson localized state \cite{al}. It occurs due to interference of the repeatedly reflected quantum matter waves in the random potential. The scenario gets changed in the presence of strong disorder potential.  Here the competition between inter-atomic interaction and disorder potentials becomes important. Specifically,  the bosons  undergo a phase transition  from super-fluid to Bose glass (BG)  due to localization in the minima of the random potential. The BG phase is a many-body insulating phase which is characterized by  finite compressibility, absence of gap in the spectrum and infinite super-fluid susceptibility \cite{bg1,bg2,bg3,bg4,bg5,NJP,Indentify}.

The effect of disorder potential  has been studied experimentally in the ultracold atomic systems either by the use of laser speckle patterns \cite{r6,r6a} or by applying incommensurable optical lattices \cite{r7,r7aa,r7bb,r7a}.  In the presence of lattice disorder  Anderson localization has directly be observed in one-dimensional configuratiion \cite{r6a,r7a}. In compositionally disorder platform Anderson localization and photonic band-tail states  have recently been confirmed experimentally \cite{p5}. Several theoretical investigations  including the finite temperature phase transition from super-fluid to insulator \cite{p1}, localized state with exponential tail \cite{p2}, disorder induced phase coherence \cite{p3} and distribution of bosons localized at the minimum of the disorder potential \cite{p4} show current interest in this direction.

A natural framework to study localization phenomena  can be provided by the information theoretic approach \cite{s1,s2,s3,bbm}. It is due to Shannon who establishes a link between physical effects and information entropy.  Shannon information entropy depends on the probability density corresponding to changes in some observables.  Quantifying observable physical effects using information entropy is useful in many applicative contexts.  For example, it is shown that the Shannon information entropy can give better information on the effect of electronic correlation in many-electron systems \cite{bkt1,bkt1a,bkt1b}.  Time evolution of the sum of entropies in position and momentum spaces  helps in understanding collapse  and revival phenomena of quantum wave packets \cite{bkt2}.

Our objective in this paper is to study effects of  quasi-disorder potentials on  the density of   matter waves in Bose-Einstein condensates (BECs). Localization of quantum matter waves results from combined effects of kinetic energy, inter-atomic interaction and disorder potentials. The disorder potential interplays with kinetic energy to determine the phase of the condensates. Non-linearity due to inter-atomic interaction further introduces complexity in the system.  An average amount of information from the event can, however, be obtained by the measure of Shannon entropy. For a continuous probability distribution $\rho(x)=|\phi(x)|^2$,  Shannon information entropy in coordinate space  is defined by
\begin{eqnarray}
S_\rho=-\int \rho(x) \,\,\text{ln}\, \rho (x)\,dx
\label{eq1}
\end{eqnarray}
with $\int \rho(x) dx=1$. Let $\psi(p)$ be the Fourier transform of $\phi(x)$, the Shannon entropy corresponding to $\gamma(p)=|\psi(p)|^2$ is given by 
\begin{eqnarray}
S_\gamma=-\int \gamma(p) \,\,\text{ln}\, \gamma(p)\,dp. 
\label{eq2}
\end{eqnarray}
In one-dimension, the values of  $S_\rho$ and $S_{\gamma}$ are shown to satisfy 
\begin{equation}
 S_\rho+S_\gamma \,\ge\, (1+\ln \pi).
\label{eq3}
\end{equation}
This inequality is often  referred to as Bialynicki-Birula-Mycielski (BBM) uncertainty relation \cite{bbm0}. It implies that if a distribution in coordinate space is localized  then  the distribution in momentum space is diffused and vice versa. We know that  disorderness augments entropy. Therefore, a larger value of {$S_\rho$} is associated with the smaller value of  $S_{\gamma}$ \cite{bbm}.

We know that localization of ultra-cold atoms in pseudo-disorder potential is affected (i) with the change of interaction strength $(\gamma)$ and/or (ii) by the variation of relative strength, wavelength and phase of the lattices constituting the pseudo-disorder potential. Understandably, atoms in BEC loaded in optical lattices are distributed in different lattice sites. With the variation of interaction the number of occupied sites gets modified. Therefore, for a particular  quasi-random potential the Shannon entropy of  density distributions in coordinate- and momentum-spaces are likely to be affected.  We see that, for a particular state with fixed width and quasi-disorder  lattice parameter, $S_\rho$ changes with the variation  of $\gamma$ and attains a minimum value at $\gamma=\gamma_m$ for maximally localized states. At the same time momentum spaces entropy tries to take lower and lower values.  However, the entropic uncertainty relation restricts the lowest values of $S_{\rho}$ and $S_\gamma$.  Therefore, $S_\rho$  along with BBM inequality can serve as the measure of localization of matter waves in Bose-Einstein condensates.

The paper is organized as follows. In Section $2$, we introduce an appropriate model and present a variational study to estimate optimum values of parameters. More specifically, we calculate energy of the system and find the values of different parameters for  energetically stable condensates. The energy variation with inter-atomic interaction suggests the existence of two different types of states in the system. In section $3$, a numerical study is presented with a view to examine the nature of the localized states. We calculate Shannon entropy  to quantify the localization in the system. Finally, we conclude by noting the main outcome of the work in section $4$.

\section{Theoretical Formulation}
The theoretical prediction on the phenomenon of localization of electro-magnetic waves in disorder potential  has recently been realized experimentally in  ultracold atomic systems by Roati $et \,\,al.$\cite{r7a}. It was performed by the use of one-dimensional quasi-periodic potential created by suerposing  two standing waves of slightly different wave lengths, namely, $\lambda_1=1032\,\,nm$ and $\lambda_2=862\,nm$ such that the value of $\lambda_r=\lambda_2/\lambda_1$ maintains  incommensurable condition. Understandably, this is a  bi-chromatic lattice potential where the effects of secondary lattice changes the periodicity and magnitude of the principal lattice in a irregular manner.  The quasi-periodic potential can be written as \cite{r7a}
\begin{eqnarray}
V_{\rm QD}(x)=\sum_{i=1}^{2}2 s_i E_i\,\cos^2(k_i\,x). 
\label{eq4}
\end{eqnarray}
Clearly, $2 s_i$ measures the strength of the quasi-periodic potential in the unit of recoil energy $E_i=\frac{\hbar^2 k_i^2}{2m}$ with $k_i=\frac{2\pi}{\lambda_i}$, the wave number. It is shown in \cite{r7a} that the value of $s_i$ must be limited by  $s_1\leq 10$ and $s_2\leq 3$. Recently, insulating phases  due to the action of weak disorder lattice have been studied by loading ultracold atoms in quasi-periodic potentials. Some differences in the properties of ultracold gases in pure-disorder and quasi-disorder potentials are observed. However, most of the properties  are shown to be  almost identical in both the cases \cite{bg}.

In view of the above, we consider a cylindrical shaped BECs of $N$ atoms in quasi-disorder potential given in Eq. (\ref{eq4}). The geometry of the condensate can be  achieved experimentally by making the frequency of the transverse  harmonic trap ($ \omega_\perp$) much greater than that of longitudinal component ($\omega_x$) of the trap such that $\omega_\perp\gg \omega_x$ \cite{Dalfavo}. The dynamics of trapped BECs is satisfactorily described by means of an effective mean-field equation, often called Gross-Pitaevskii equation (GPE). The  GPE in quasi-one-dimension(Q1D) is given by
\begin{eqnarray}
i\hbar\dfrac{\partial \psi}{\partial t}&=&-\dfrac{\hbar^2}{2m}\dfrac{\partial^2 \psi}{\partial x^2}+\frac{1}{2}m \omega_x^2 x^2 \psi+ V_{\rm QD}(x)\psi\nonumber\\&+&2a_sN\hbar \omega_\perp |\psi|^2 \psi
\label{eq5}
\end{eqnarray}
with $\int|\psi|^2\,dx=1$. Here $a_s$  stands for the atomic scattering length. Understandably,  Eq.(\ref{eq5}) models the BECs in quasi-disorder potentials.  Rescaling  length, time and energy  in the units of $a_\perp=\sqrt{\hbar/(m \omega_\bot)}$, $\omega^{-1}_\bot$ and $\hbar\omega_\bot$ we get 
\begin{equation}
 i\dfrac{\partial \phi}{\partial t}=-\dfrac{1}{2 }\dfrac{\partial^2 \phi}{\partial x^2}+\frac{1}{2} \lambda_x^2 x^2 \phi+ V_{\rm QD}(x)\phi+\gamma|\phi|^2 \phi
 \label{eq6}
\end{equation}
 with
\begin{eqnarray}
V_{\rm QD}(x)=\sum_{i=1}^{2}{s_i k_i^2}\,\cos^2\left({k_i x}\right),
\label{eq7}
\end{eqnarray}
where  $\lambda_x=\omega_x/\omega_\perp$, $\gamma=2 a_s N /({\cal N} a_\perp)$ and $\phi=\sqrt{a_{\perp}}\psi$  with the norm ${\cal N}=\int |\phi(x,t)|^2\,dx$. We introduce, for convenience, $k_1=k_p$ and $k_2=k_s$ and represent $V_p=s_1k_p^2$ and $V_s=s_2 k_s^2$ as the strengths of primary and secondary lattices respectively. 

Different approximation methods are usually used to solve the GP equation in (\ref{eq5}) \cite{antun1,antun2,antun3,antun4}. One of the widely used methods to treat the problem  analytically is the variational approach \cite{golam2,golam3,golam4,Anderson}. To begin with we restate the initial-boundary value problem in Eq.(\ref{eq6}) by writing the Gross-Pitaevskii energy functional
\begin{eqnarray}
E[\phi,\phi^*]\!\!=\!\!\!\int_{-\infty}^{+\infty}\!\!\!\left( \frac{1}{2} \left|\phi_x\right|^2\!\!
+\!\!V_{\text{D}} \left|\phi\right|^2\! +\! \frac{1}{2} \lambda_x^2 x^2 \left|\phi\right|^2 \!+\!\frac{1}{2}\gamma \left|\phi\right|^4\!\! \right)\!\!dx.
\label{eq8}
\end{eqnarray}
Here, the first and last terms represent  respectively  kinetic  and interaction energies while the second and third terms give contributions of quasi-periodic potential and harmonic trap.
In order to understand the dependence of localized states on different parameters of the system, we consider 
\begin{eqnarray}
\phi(x)=\sqrt{\frac{\cal N }{\sqrt{\pi} a}}\,\, e^{-x^2/2 a^2 },
\label{eq9}
\end{eqnarray}
as a trial solution.  Here  $a$ is the variational parameter. This is a reasonable assumption  of localized states since in the absence of inter-atomic interaction and $V_{QD}=0$, the  solution of Eq.(\ref{eq6}) is a Gaussian function. The inter-atomic interaction changes only size of the condensate cloud. We demand that this change will be reflected in the change of variational parameter.  To determine values of the parameters for $V_{QD}\ne0$ and $\gamma\ne0$ we insert Eq.(\ref{eq9}) in Eq. (\ref{eq8}) and calculate
\begin{eqnarray}
\frac{E}{\cal N}\!\!\! &=&\!\!\frac{1}{4 a^2}+ \frac{1}{4} \lambda_x^2 a^2\!\!+\!\! \frac{{\cal N} \gamma}{2  \sqrt{2 \pi}a}
\!+\frac{1}{2} V_p \left(1+e^{-{k_p^2 a^2 }}\right)\nonumber\\&+&\frac{1}{2} V_s\left(1+ e^{-{k_s^2 a^2}}\right)\!\!.
\label{eq10}
\end{eqnarray}
We know that the necessary condition for optimization i.e., $\partial E/\partial a = 0 $ allows us to determine the effective values of $a$. Let $a_0$  be the value of  $a$ which satisfies the sufficient condition $\partial^2 E/\partial a^2 |_{a=a_0}>0$ and, thus permits energetically stable condensates.  However, it is an interesting curiosity to find effective energy for fixed quasi-periodic lattice parameters but different interaction strengths or vice versa. We expect that the  interplay between the quasi-periodic and nonlinear potentials in creating localized/extended state can change the shape of the effective potential.

\begin{figure}[h!]
\begin{center}
\includegraphics[width=7cm, height=5cm]{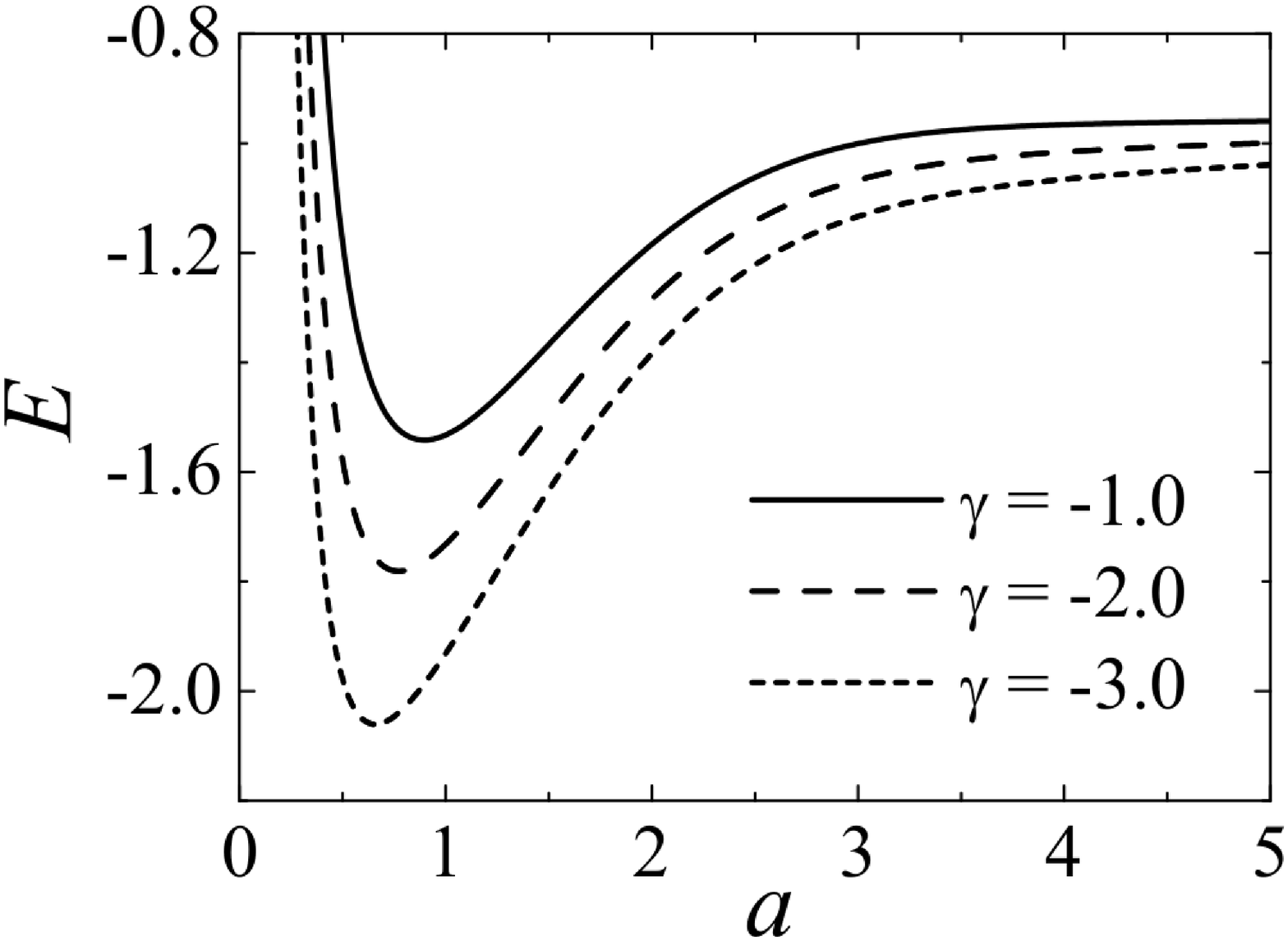}
\includegraphics[width=7cm, height=5cm]{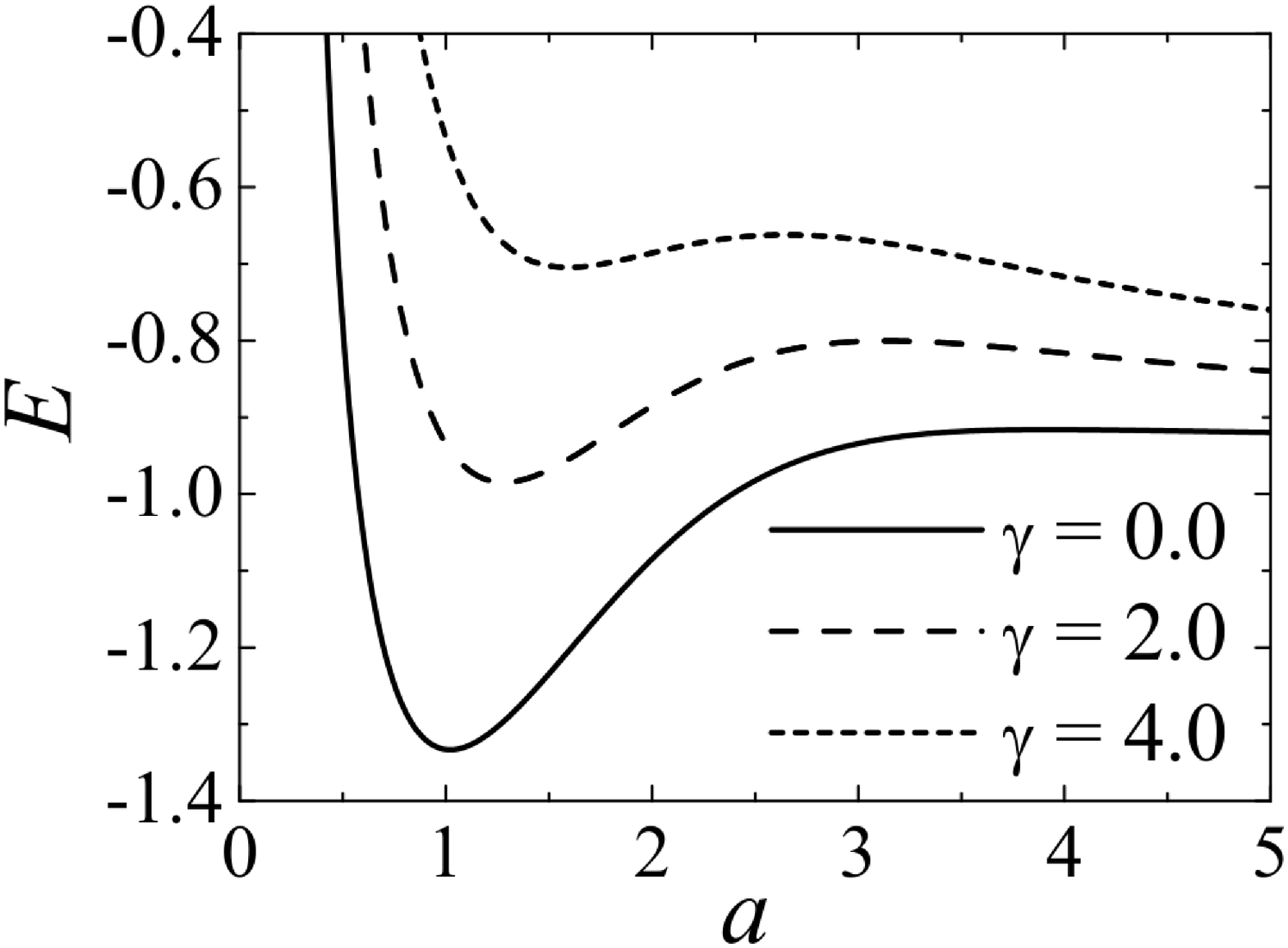}
\caption {Top panel. It shows energy ($E$) as a function of $a$ for  $\gamma < 0$.   Bottom panel.  It gives $E$ versus $a$ for $\gamma\geq 0$. In both the panels we consider $s_1= -6.0\,$, $s_2=0.9\,$, $\lambda_p=10$, $\lambda_s=0.835\lambda_p$, ${\cal N}=1$. Also we assume that the trap is sufficiently flat and thus the effects of  $\lambda_x$ can be negligible at the central region of the harmonic confined. }
\end{center}
\label{fig1}
\end{figure}

With a view to  examine the interplay between nonlinear and quasi-disorder potentials, we choose different parameters within the experimental limit.  More explicitly, we have chosen to work with
$\lambda_p=10$ and  $\lambda_s=8.38$ which give actual wavelengths  of the bi-chromatic lattices in the unit of  $a_\perp \approx 1 \mu m$. The ratio of secondary and primary lattices thus consistents with the experinemtal value $\lambda_r=0.835$ \cite{r7a}. The choice of $\lambda_p$ is not unique. Indeed, one can work with different values of $\lambda_p$ keeping $\lambda_r$ at $0.835$ \cite{ref46}. Strengths of the lattices  can be calculated from $V_p=4\pi^2  s_1/\lambda_p^2$ and $V_s=4\pi^2 s_2/\lambda_s^2$ for $s_1\leq 10$ and $s_2\leq 3$. Thus fixing the values of lattice parameters, we plot in Fig. $1$ effective energy $E$ as a function of $a$ for different interaction strengths. Specifically, top and bottom panels give $E$ versus $a$ for attractive and repulsive interactions respectively.  The existence of negative minima at certain values of $a$ indicates that a BEC can hold stable bound states in both repulsive and attractive cases. However, the value of $a$ of a stable condensate decreases as attractive interaction increases while a state becomes extended as the repulsive interaction increases.

It is worth mentioning that negative minimum appears in the effective potential if the  quasi-periodic pattern arises from two lattices which are out-of-phase. However, one can get localized states by superposing two in-phase lattices. In this case  minimum energy becomes positive and localization takes place only if $|s_2|/|s_1|\ll 1$ \cite{ref46}.

\section{Information Entropy of BEC matter waves} 
We know that different types of localized states are observed in the BECs  in the presence of periodic and pseudo-random potentials. The interplay between  lattice potential and nonlinear interaction gives rise spatially localized waves \cite{ref47}. We have seen that analytical approximation method can help in understanding the interplay between different parameter of the system and allow us to estimate width/size of the condensate distribution.  However, exact shape or nature of the distribution plays a crucial role in the entropic based approach. In this context we  note that the Gaussian type states having different widths obtained from the energy optimization condition for different $\gamma$ cannot detect the variation of entropic uncertainty in the Shannon formalism. This leads us to consider a numerical based approach to get more detail description of the density profile.
 
\subsection{Information entropy  in periodic potentials}
We consider both non-interacting  and  interacting  BECs in periodic potentials.  Depending on the type of interaction (attractive and repulsive) the system can be responsible for different types of localized states e.g., bright and gap solitons,  the physical origin of which are quite different. To begin with we perform the numerical simulation \cite{ref48} of Eq.({\ref{eq6}}) with potential given in Eq.(\ref{eq7}) for $V_s=0$. The result of which is displayed in Fig. 2.
\begin{figure}[h!]
\begin{center}
\includegraphics[width=7cm, height=5cm]{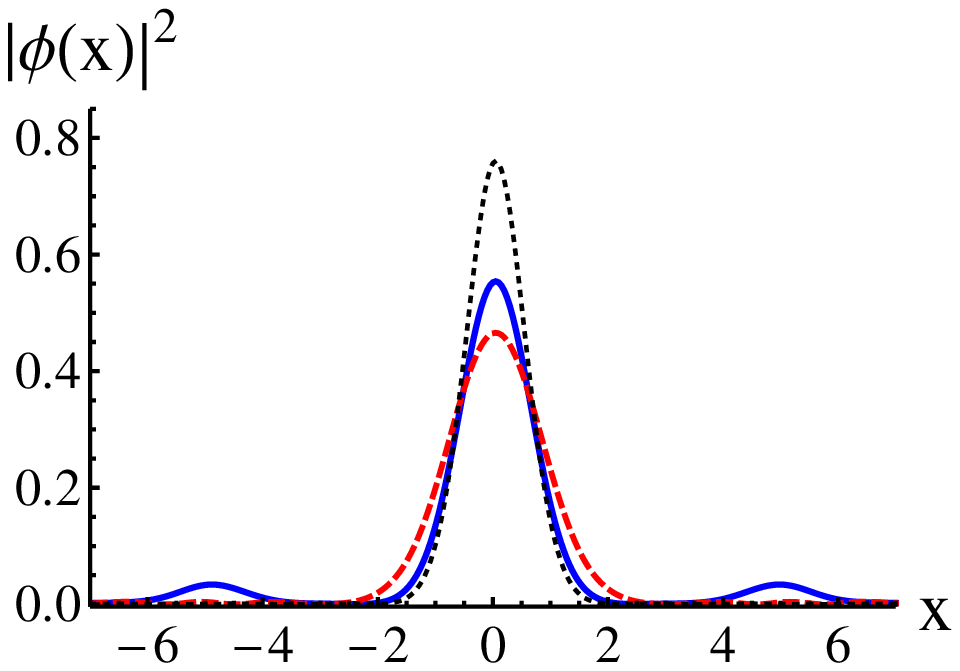}
\includegraphics[width=7cm, height=5cm]{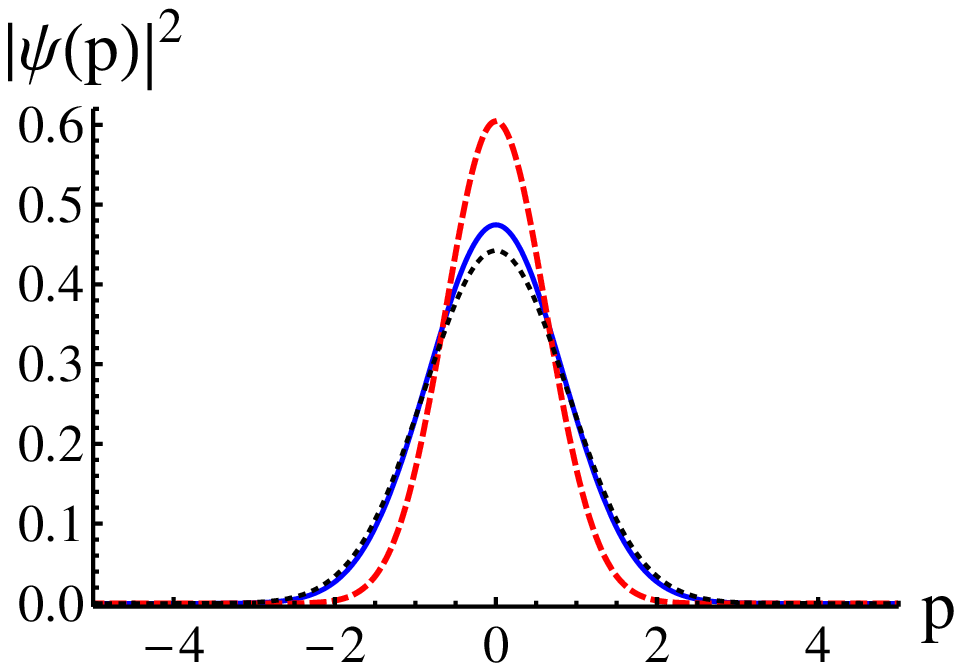}
\caption { Color(online) Top panel shows density distribution ($|\phi|^2$) as a function of $x$ for different interaction strength $\gamma$ at $V_p=-6.0$, $V_s=0$, $\lambda_p=10$ and $\lambda_r=0.835$  in co-ordinate spaces. A similar variation in momentum spaces fitted with Gaussian function is shown in the bottom panel. Solid, dashed and dotted curve give density distributions for $\gamma=0, 2$ and $-1$.}
\end{center}
\label{fig2}
\end{figure}	
Particularly, it gives density distributions both in co-ordinate (top panel) and momentum (bottom panel) - spaces  for  $V_p=-6$ and $\lambda_p=10$.  Top panel shows that the peak (spatial distribution) of density distribution of a bound state decreases (broadens) as the interaction changes from attractive to repulsive. As expected, the momentum space shows the opposite trends. Entropies of different states are shown in Table $1$.  We see that the $S_\rho$ gradually decreases (increases)  with the increase of attractive(repulsive) interaction. 
 An opposite trend in entropy  variation is found in the momentum-space distribution. Therefore, a clear indication is that  the spatial extension of BEC matter waves in co-ordinate space reduces if the interaction is tuned from  repulsive to attractive. The maximum localization is, however, limited by the entropic uncertainty in Eq. (\ref{eq3}). This entropic uncertainty led us to demand that the sum of $S_\rho+S_\gamma$ tends towards the lowest value for the most feasible state. 

\begin{table}[h!]
\centering
\begin{tabular}{|c|c|c|c|c|c|c|}
\hline
$\gamma$&$a$&$S_\rho$&$S_\gamma$&$S_\rho+S_\gamma$\\
\hline
-2.0&0.5972&0.7928&1.4451&2.2379\\
\hline
-1.0&0.6556&1.0503&1.2756&2.3259\\
\hline
0.0&0.7127& 1.5743&1.2350&2.8092\\
\hline
1.0&0.7681&1.3972&1.0906&2.4878\\
\hline
2&0.8219&1.5447&1.0704&2.6151\\
\hline
\end{tabular}
\caption{Shannon entropies and width of different states obtained by the variation of $\gamma$ of BECs in a periodic potential.}
\end{table}
Looking carefully Table $1$ we see that the width of  a state reduces gradually with the tuning of inter-atomic interaction from repulsive to  attractive through $\gamma=0$. The  entropic uncertainty obtained from the direct numerical simulation of Eq.(\ref{eq6}) gradually increases except for weakly interacting case. More specifically, entropic uncertainty for $\gamma=0$ becomes larger than those of the interacting cases. This can be understood by noting that the  central peak of the distribution for $\gamma=0$ decreases due to appearance of new smaller peaks on either side of central maximum resulting an effective broadening of the state. It may be remembered that the localized matter waves for $\gamma<0$ and $\gamma>0$ can be identified as bright and gap solitons respectively \cite{ref47}. For brevity we say that the high information entropy at $\gamma=0$  causes spatial broadening of the profile. 

\subsection{Information entropy in quasi-periodic potentials}
We have noted that two lattices with different wave-numbers, strengths and phases  are superposed to create a quasi-periodic potential.  It is seen that the effective energy of BEC can either be positive or negative  depending on lattice parameter (Fig. $1$).  The BEC in quasi-periodic potential with positive effective energy  exhibits Anderson type localization in the $\gamma\rightarrow 0$ limit due to interplay between kinetic energy and optical lattices. An interacting BEC with negative effective energy, on the other hand, can support localized states. In this case, the interplay between pseudo-random and nonlinear potentials in conjunction with kinetic energy comes into play to control  nature of the states having negative effective energy.   

To quantify the localization of those states  in terms  of information entropy we calculate density profile from numerical simulation of the Eq.(\ref{eq6})  with the potential given in Eq.(\ref{eq7}). The result on the density distribution $|\phi(x)|^2$ is displayed in Fig.$3$. It is seen that the localization of  states supported by quasi-disorder lattice becomes narrower in coordinate space as the attractive interaction  increases. As expected momentum space distribution exhibits opposite trends. We know that both momentum and coordinate space distributions,  in principle,  tend to be more localized to attain  minimum entropy. The minimum entropic uncertainty relation, however, limits the lowest value of entropy for localization. We find that the physically acceptable most localized state attains lowest value of the entropic uncertainty condition. The listed data in Table $2$ clearly implies the fact that the state with $\gamma=-1$ should be more physically acceptable localized states. Interestingly enough, the co-ordinate space distribution shows more localization with the increase of attractive interaction resulting smaller entropy. In that case sum of entropies increases due to faster increase of momentum-space entropies leading to less feasible states.
\begin{figure}[h!]
\begin{center}
\includegraphics[width=7cm, height=5cm]{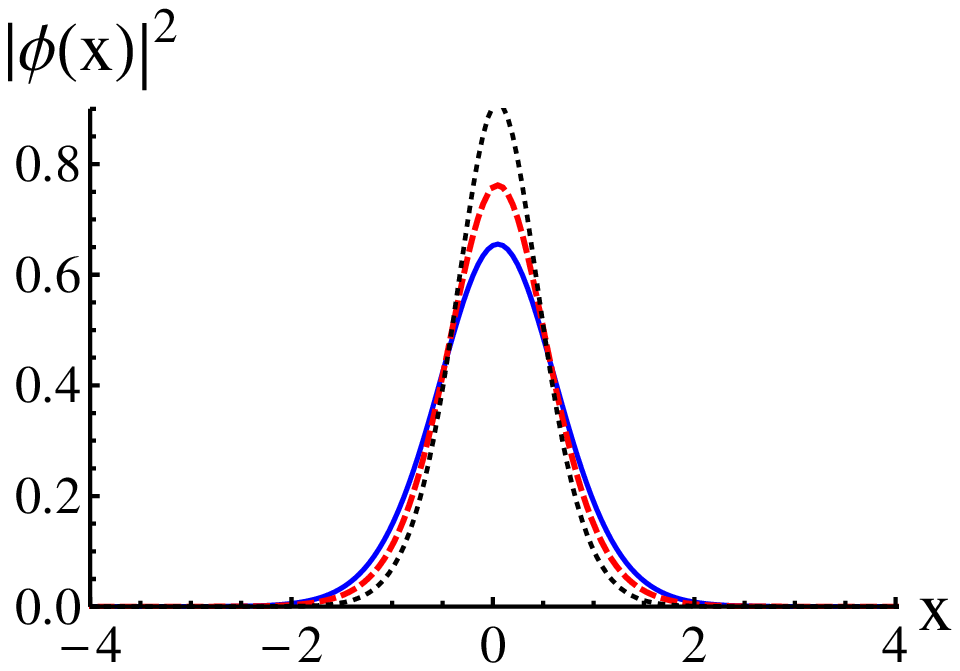}
\includegraphics[width=7cm, height=5cm]{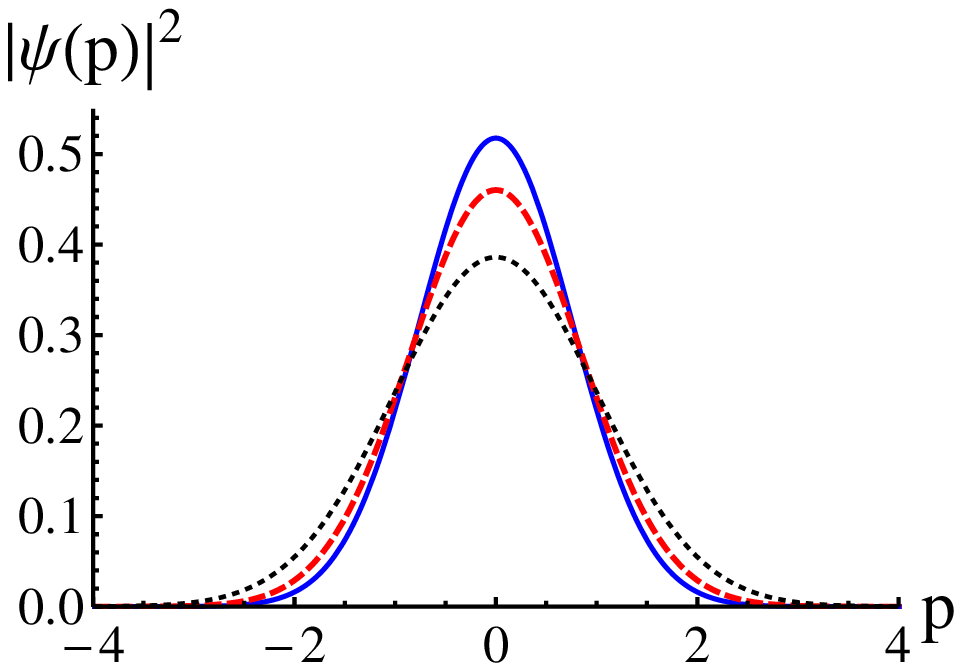}
\caption { Color(online) Top panel gives density distribution ($|\phi(x)|^2$) in coordinate space  for attractive inter-atomic interaction.
Bottom panel  displays density distribution ($|\psi(p)|^2$) obtained from Gaussian fit in momentum-space. In both the panels solid, dashed and dotted curve represent states for  different  values of attractive interaction strength, namely, $\gamma=-1,-2$ and $-3$ respectively with $V_p=-6.0$ and $V_s=0.9$.}
\end{center}
\label{fig3}
\end{figure}
\begin{table}[h!]
\centering
\begin{tabular}{|c|c|c|c|c|c|c|}
\hline
$\gamma$&$a$&$S_\rho$&$S_\gamma$&$S_\rho+S_\gamma$\\
\hline
-1.0&0.8962&1.0002&1.1675&2.1687\\
\hline
-2.0&0.77405&0.8096&1.3788&2.1884\\
\hline
-3.0&0.6587&0.6987&1.5050&2.203\\
\hline
-4.0&0.5750&0.5124& 1.7106&2.2229\\
\hline
-5.0&0.4715&0.4072&1.8274&2.2347\\
\hline
\end{tabular}
\caption{Shannon entropies and width of different states in BECs in quasi-periodic potential for different values of $\gamma$. Other parameters are kept same as those used in Fig. 3.}
\end{table}
\begin{figure}[h!]
\begin{center}
\includegraphics[width=7cm, height=5cm]{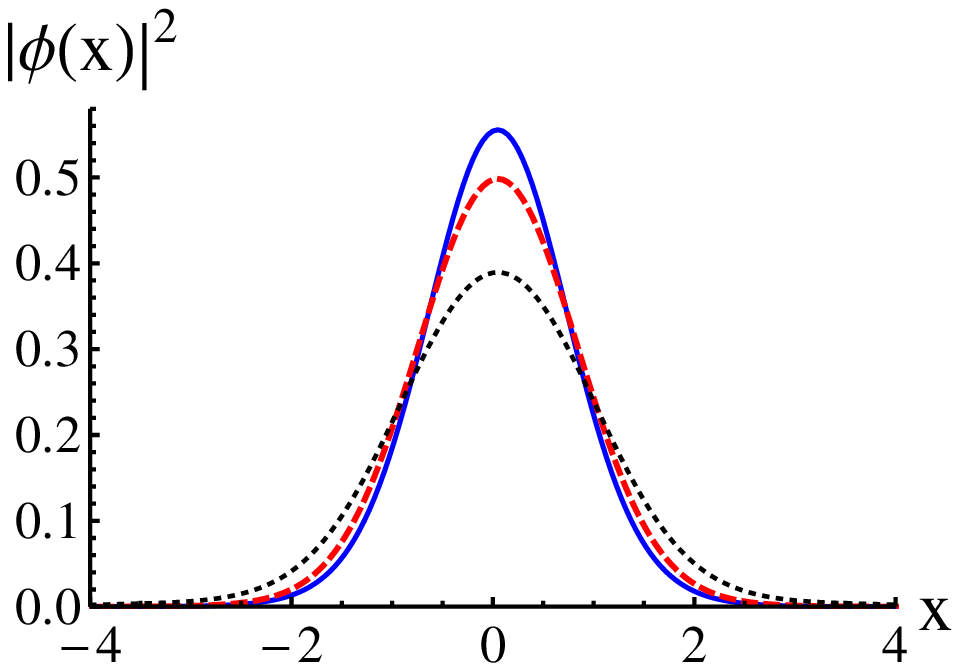}	
\includegraphics[width=7cm, height=5cm]{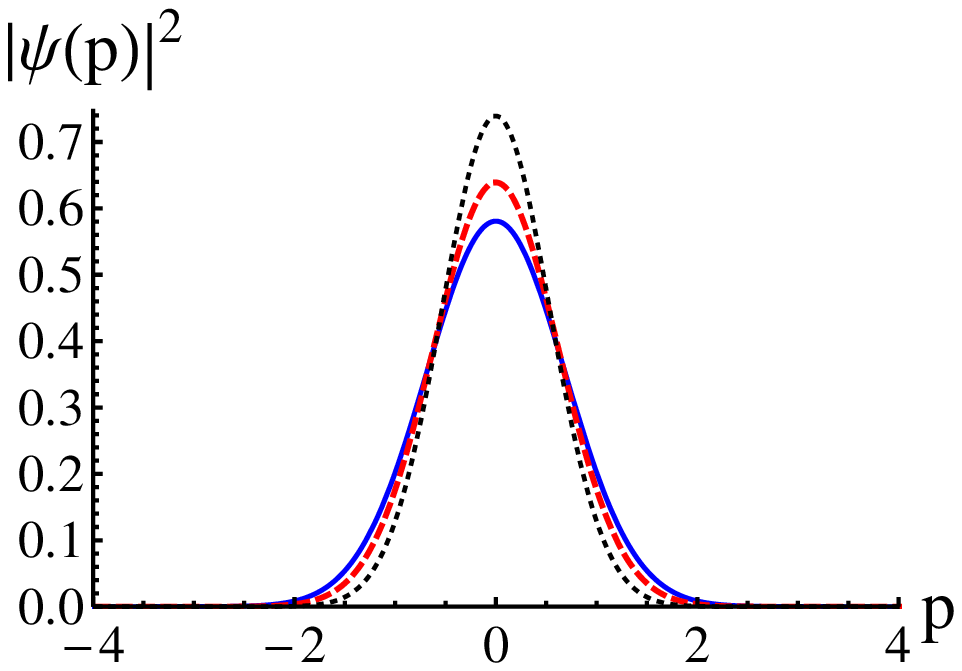}
\caption {Color(online) Top Panel : Density distribution ($|\phi(x)|^2$) in coordinate space.
Bottom panel :  The variation of ($|\psi(p)|^2$) obtained from Gaussian fit in momentum-space for the values of different parameters similar to those used in the top panel. In both the panel solid, dashed and dotted curve represent states for  different  values of repulsive interaction strength, namely, $\gamma=0,1$ and $3$ respectively with $V_p=-6.0$, $V_s=0.9$ and $\lambda_p=10$.}

\end{center}
\label{fig4}
\end{figure}
\begin{table}[h!]
\centering
\begin{tabular}{|c|c|c|c|c|c|c|}
\hline
$\gamma$&$a$&$S_\rho$&$S_\gamma$&$S_\rho+S_\gamma$\\
\hline
0.0&1.02126&1.1420&1.0308&2.1728\\
\hline
1.0&1.1487&1.2524&0.9257&2.1807\\
\hline
2.0&1.2813&1.3931&0.8039&2.1970\\
\hline
3.0&1.4249&1.6143&0.7632&2.3775\\
\hline
4.0&1.5938&2.7286&0.9941&3.7228\\
\hline
\end{tabular}
\caption{Shannon entropy and entropic sum of different states of  BECs in quasi-periodic potential for different values of repulsive interaction. Other parameter are same as those used in Fig. $4$.}
\end{table}
\begin{figure}[h!]
\begin{center}
\includegraphics[width=7cm, height=5cm]{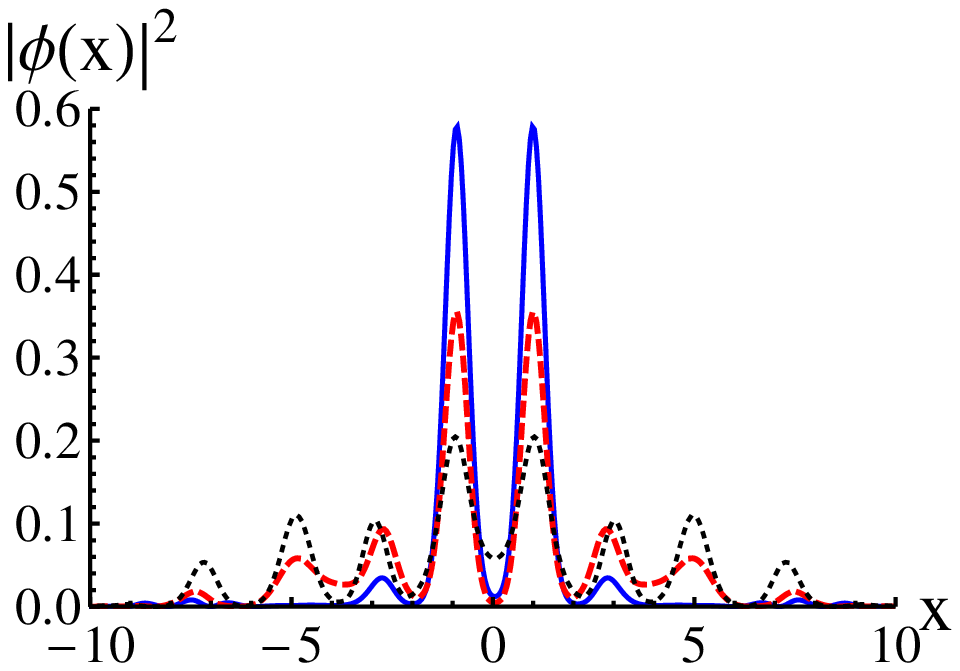}	
\vskip 0.25cm
\includegraphics[width=7cm, height=5cm]{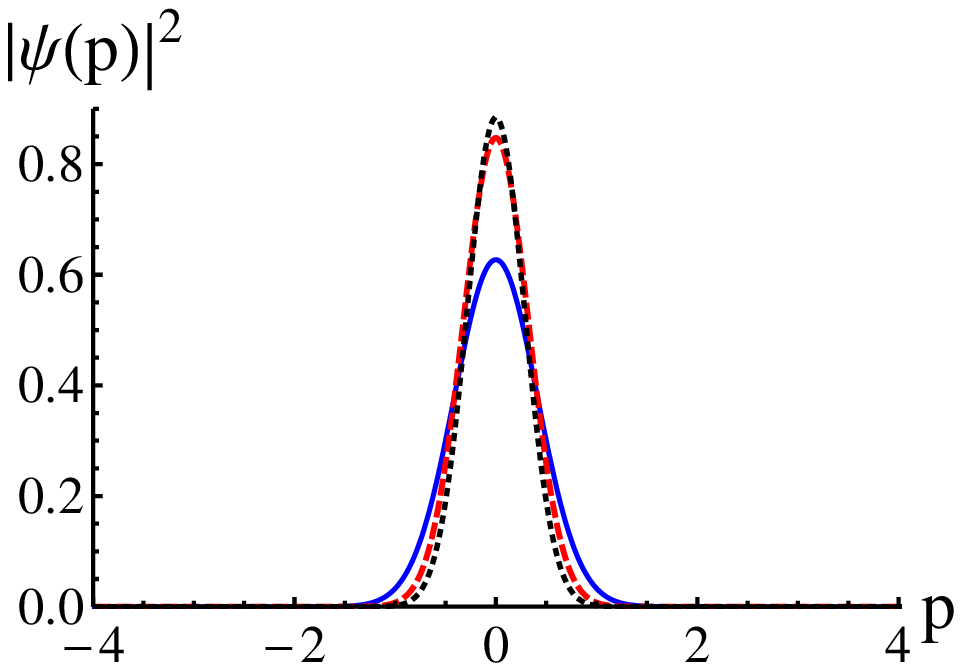}
\caption {Color(online). Top Panel gives density distribution ($|\phi(x)|^2$). Bottom panel shows density distribution ($|\psi(p)|^2$) in momentum space for same values of $\gamma$ as those used in the Bottom panel. In both the panels solid, dashed and dotted curves represent the states for interaction strengths $\gamma=0,1.5,2$ with $V_p=2.0$ and $V_s=0.4$, $\lambda_p=4$. Momentum space distributions displayed here are obtained from Gaussian fit to visualize localization effect clearly.}
\end{center}
\label{fig5}
\end{figure}

It is interesting to note that the localized state with negative energy  shows different behaviour in the presence of repulsive interaction due to quasi-disorder lattices. Here the quasi-disorder potential interplays with kinetic energy  in the mechanism of localization / delocalization phenomena. We notice that the distribution of density profile is maximally localized  in the coordinate space  and diffusely distributed in the momentum-spaces with $\gamma=0$ (see Fig. $4$).  More specifically, with the increase of $\gamma$ the central narrow peak diminishes which results weak localization in coordinate spaces and strong localization in momentum space.  The entropies  in coordinate and momentum spaces show opposite trends for $\gamma>1$. However, the sum of the two entropies increases for $\gamma>0$. The most feasible state is the one which has minimum entropic uncertainty.

\begin{figure}[h!]
\centerline{
\includegraphics[width=7cm, height=5.5cm]{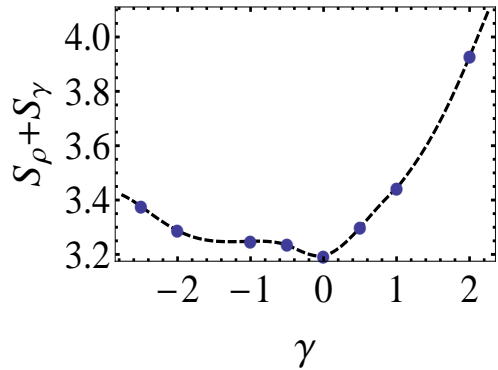}}
\caption {Color(online) Entropic uncertainty $S=S_\rho+S_\gamma$ of a particular state with the variation of  interaction strength $\gamma$.}
\label{fig6}
\end{figure}

A localized state with positive energy can be created in a BEC with repulsive inter-atomic interaction in the presence of a bi-chromatic lattice which results from the superposition of  in-phase optical lattices. In this case, the central peak gets localized in the two nearby sites of the central lattice maximum (Fig. $5$). Understandably, physical origin of this type of localised state is different from that of a bound state. Here localization takes place due to multiple reflections from lattice disorder and there is no role of interaction (see curve for $\gamma=0$ in Fig. $5$). We have seen that the entropic uncertainty is minimum for $\gamma=0$  and it increases for $\gamma \neq0$. We demand that stronger localization of the state causes entropy to decrease. However, maximally localized state corresponds to that   which  gives minimum  entropic uncertainty (Fig. \ref{fig6}).

\section{Conclusion}
We have investigated the phenomenon of spatial localization of  Bose-Einstein condensates in periodic and quasi-periodic  potentials separately within the framework of Gross-Pitaevskii equation. In particular, we have studied the role of  nonlinear interaction and quasi-disorder potential  in creating   localized or extended bound states. Our investigations on the existence of different types of states for interacting and weakly-interacting BECs are based on both analytical and numerical approaches. It is found that the BECs can support  a family of states, the  spatial extension of which can be varied by changing the interaction strength and with the variation of relative strength, wave-number and phases of the constituents of bi-chromatic lattices.

We have quantified localization of a state by the measure of Shannon information entropy. It is seen that the information entropy of Shannon in position-space ($S_\rho$) is a useful physical quantity to measure localization of any kind of states. Smaller value of $S_\rho$ infers that the localization is more in coordinate space while converse is true for the entropy $S_\gamma$ in momentum space. Both the entropies try to get smaller and smaller values to produce more and more localized density profile. Their lower values are, however, restricted by the entropic uncertainty relation. More specifically, uncertainty  sum attains lowest value for the most feasible localized state.

We have seen that with the increase of attractive (repulsive) interaction localization of BEC distribution gradually increases (decreases) without any upper (lower) bound in  coordinate-space. The  entropy based approach can allow us to identify a critical value of interaction strength or lattice parameter for maximally localized  BEC wave-packets.

We have studied information entropy of  non-interacting BECs in quasi-disorder and periodic potentials. Our study shows that the entropic uncertainty for the bound state embedded in quasi-disorder potential is smaller than that in periodic potential. The non-interacting BECs in presence of in-phase bi-chromatic lattice can also support state with positive energy. We have verified that the entropic uncertainty of all possible  states in the presence of  interaction either attractive or repulsive becomes larger than that of the state in non-interacting BECs. This investigation clearly reveals the fact that disordering in the lattice is responsible for localization. 
\vspace{0.5cm} 
\subsection*{Acknowledgements}
 We acknowledge the hospitality from all teaching members of the  Department of Physics, B. B. College, Asanasol, West Bengal, India.


\begin{thebibliography}{99}
\bibitem{ref48} The numerical simulation is performed using  Split-step Fourier method with $\Delta x=16 \pi/1024$ and $\Delta t=0.001$.
\end{thebibliography}
\end{document}